\documentclass[prd,nofootinbib]{revtex4}
\usepackage{amsmath,amssymb}
\usepackage{graphicx}
\usepackage{rotating}
\usepackage{bm}

\usepackage{color}

\newcommand{\eosa}{EOS I}
\newcommand{\eosb}{EOS II}

\begin{document}

\author{Robert R. Caldwell${}^{1}$}
\email{Robert.R.Caldwell@Dartmouth.edu}
\author{Michael Doran${}^{1,2}$}
\email{M.Doran@gmx.de}
\affiliation{${}^1\,$Department of Physics \& Astronomy,
        HB 6127 Wilder Laboratory,
        Dartmouth College,
        Hanover, NH 03755, USA\\
${}^2\,$Institut f\"ur  Theoretische Physik, Philosophenweg 16, 69120 Heidelberg, Germany}
\title{Dark-Energy Evolution Across the Cosmological-Constant Boundary}

\begin{abstract}

We explore the properties of dark energy models for which the
equation-of-state, $w$, defined as the ratio of pressure to energy density,
crosses the cosmological-constant boundary $w=-1$. We adopt an empirical
approach, treating the dark energy as an uncoupled fluid or a generalized
scalar field. We describe the requirements for a viable model, in terms of the
equation-of-state and sound speed. A generalized scalar field cannot safely
traverse $w=-1$, although a pair of scalars with $w >-1$ and $w < -1$ will
work. A fluid description with a well-defined sound speed can also cross the
boundary. Contrary to expectations, such a crossing model does not
instantaneously resemble a cosmological constant at the moment $w=-1$ since the
density and pressure perturbations do not necessarily vanish. But because a
dark energy with $w < -1$ dominates only at very late times, and because the
dark energy is not generally prone to gravitational clustering, then crossing
the cosmological-constant boundary leaves no distinct imprint.

\end{abstract}

\maketitle

 
Numerous observations and experiments indicate that our universe has low
matter-density, negligible spatial curvature, and is currently undergoing
accelerated cosmic expansion
\cite{Spergel:2003cb,Readhead:2004gy,Goldstein:2002gf,Rebolo:2004vp,Riess:2004nr,Tegmark:2003ud,Hawkins:2002sg}.
The remarkable implication is that approximately two-thirds of the cosmic
energy is due to some form of as-yet unidentified dark energy. While the
leading interpretation is that the dark energy is due to a cosmological
constant, the physical origin of such a constant remains a mystery, and it is
widely regarded as a placeholder until a deeper understanding of the dark
energy can be established. 

In an effort to characterize the nature of the dark energy, attention has
focused on its presumed equation-of-state, $w$, defined as the ratio of its
mean pressure to energy density, $w \equiv p/\rho$. A cosmological constant
corresponds to $w=-1$. Another conjecture is that the dark energy is due to
quintessence, a dynamical, ultra-light scalar field with negative pressure, for
which $w > -1$ \cite{Wetterich:fm,Ratra:1987rm,Caldwell:1997ii}.  A separate
class of models with $w < -1$, representing an exotic field or perhaps new
gravitational phenomena, are also under investigation ({\it e.g.}
\cite{Caldwell:1999ew}). Extensive analysis of the cosmological predictions for
all these cases finds that the current data favors dark energy models with an
equation-of-state in the vicinity of $w=-1$, straddling the
cosmological-constant boundary.

If indeed dark energy with $w<-1$ is within the realm of possibilities, then it
would seem inevitable to inquire about a transition from $w > -1$ to $w < -1$.
This question has been taken up recently \cite{Vikman:2004dc,Hu:2004kh}; here
we contribute our results and perspective on the issue. This article examines
possible mechanisms by which dark energy can cross $w=-1$. We assume that
Einstein's gravitation is valid and that the dark energy system interacts only
gravitationally with the rest of the world --- all of our ignorance is captured
in $w$. The sticking point is the stability of such a ``crossing component,"
which brings into question the physics of the dark energy.

Let's start from the observations: measurements of luminosity distances based on
type 1a supernovae and other phenomena imply a trajectory $a(t)$ for the
expansion scale factor in our homogeneous, isotropic universe. General
relativity connects this expansion history with the matter and energy sources,
in our (approximately) geometrically flat universe, which leads us to infer the
existence of a dark energy. Making the simplest assumptions about the nature of
this unknown substance, we hypothesize that it can be described as an ideal
fluid with a mean energy density and pressure. We can parametrize the dark
energy evolution with the present-day abundance $1-\Omega_{m}$ and the
equation-of-state trajectory $w(\tau)$ as well as a sound speed for the response
of small fluctuations. This procedure is sufficient to allow us to pursue
classical tests of cosmology.

A proper comparison of dark energy model predictions with the observed cosmic
microwave background anisotropy and galaxy clustering requires that we treat
the inhomogeneities in the dark energy. We adopt the conventions and notation
of \cite{Ma:1995ey}, so that we may describe the fluid perturbations according
to the conformal-Newtonian gauge equations
\begin{eqnarray}
\dot\delta&=&-(1+w)(\theta - 3\dot\phi) - 3 {\dot a \over a}
(\delta p - w \delta \rho)/\rho \cr
\dot\theta&=&-{\dot a \over a}(1 - 3 w)\theta - {\dot w \over 1+w}\theta
+{\delta p/\delta\rho \over 1+w} k^2\delta + k^2(\psi-\sigma). 
\end{eqnarray}
The standard fluid perturbation equations appear to grow singular because the
terms proportional to $(1+w)^{-1}$ diverge  in the case of a crossing. However,
the physically meaningful source of momentum transfer in the standard
perturbation equations is $\rho(1+w)\theta$, not $\theta$ alone, so that by
defining the fractional momentum density transfer ${\cal V} \equiv (1+w)\theta$
we obtain 
\begin{eqnarray}
\dot\delta &=& -{\cal V} +3(1+w)\dot \phi  - 3 {\dot a \over a}\left(
\delta p  - w \delta\rho\right)/\rho \cr 
\dot{\cal V} &=& -{\dot a \over a}(1 - 3 w){\cal V}  +k^2 {\delta p  / \rho }
+k^2 (1+w)(\psi-\sigma).   \label{eqn::evolution}
\end{eqnarray}
In the synchronous gauge we obtain
\begin{eqnarray}
\dot\delta &=& -{\cal V} -(1+w)\frac{1}{2} \dot h  - 3 {\dot a \over a}\left(
\delta p  - w \delta\rho\right)/\rho \cr 
\dot{\cal V} &=& -{\dot a \over a}(1 - 3 w){\cal V}  +k^2 {\delta p  / \rho }
-k^2 (1+w)\sigma .   \label{eqn::evolution2}
\end{eqnarray}
With the fluid perturbation equations recast in the above forms, we see that 
the response of a dark energy density perturbation to an external gravitational
field flips sign at crossing. That is, gravitational instability becomes an
antigravitational instability. On small scales the effects of shear, which
typically damp perturbation growth, are reversed. However, there is nothing in
the equations to suggest that the fluid perturbations should vanish at the
instant $w=-1$, as we might expect if the dark energy instantaneously looks like
a cosmological constant. Without a model of the pressure fluctuations and shear
this system of equations is incomplete. 

We can close the system of equations and follow the evolution by specifying a
relation $\delta p = v^2[\tau,k]\delta\rho$ and a function $\sigma[\tau,k]$,
but there are many factors to consider. A canonical scalar field has $v^2=1$ on
small scales; on scales approaching the horizon the relation becomes
gauge-dependent. Models of a generalized scalar, such as k-essence
\cite{Armendariz-Picon:2000ah}, can have a variable $v^2$. And if $v^2 \ll 1$
within the horizon then dark energy can cluster
\cite{Erickson:2001bq,DeDeo:2003te,Bean:2003fb,Weller:2003hw}. Note that the
adiabatic sound speed $c_s^2|_{(adiab.)} \equiv \dot {p} / \dot {\rho} = -d \ln
(1+w) / 3 d \ln a$ is a red herring, as it does not actually describe the
propagation of spatial inhomogeneities in the dark energy fluid. The
inhomogeneous fluctuations generally have more degrees of freedom than the
homogeneous background. Of course, whenever $v^2 \neq c_s^2|_{(adiab.)}$, the
rate of entropy generation $\Gamma = (v^2 - c_s^2|_{(adiab.)})\delta / w$ is
non-vanishing. However, let us proceed to construct an admittedly naive,
synthetic model with $\sigma = 0$, $v^2 = 1$ on sub-horizon scales to prevent
clustering like dark matter, and $v^2 \to w$ on super-horizon scales so that
the unevolved dark energy perturbations resemble the background. (We are aware
of the many myths surrounding super-horizon modes \cite{Press:1980}.) Our naive
model has $v^2[\tau,k] = w(a)\exp(-x) + \mu^2  [ 1 - \exp(-x)]$ applied in the
conformal-Newtonian gauge, and where $\mu^2=1$ is the small scale speed of
sound and $x \equiv k \tau$. It will also be interesting to allow $\mu^2=0$.
For further variety we propose a second synthetic model with $v^2=1$ on all
scales, again in the conformal-Newtonian gauge.

The generalized scalar field, employed in k-essence and phantom models, might
also serve to describe a dark energy component which crosses $w=-1$. In this
scheme, the scalar field Lagrangian originates as a nonlinear function of
gradient and field, 
\begin{equation}
S_{d} = \int d^4 x \sqrt{-g} F(X,\varphi).
\end{equation}
Here $X \equiv \frac{1}{2}(\partial_\mu \varphi)(\partial^\mu \varphi)$ and a
canonical scalar field is simply given by $F=X + V(\varphi)$. The
spatially-uniform energy density and pressure are $\rho =
F-2XF_{,X}$ and $p=-F$. For a field with only linear dependence on $X$,
$F=K(\varphi)X+V(\varphi)$ there are two immediate consequences. First, the
system can be transformed into a canonical scalar field with positive kinetic
energy ($w > -1$) or negative kinetic energy ($w < -1$) by a field
redefinition, where we note that the equation-of-state is given by $w = F/(2 X
F_{,X} - F)$. Second, such a system leads to well-behaved pressure fluctuations
\cite{Erickson:2001bq,DeDeo:2003te,Bean:2003fb},
\begin{equation}
\label{eqn::presseqn}
\delta p = \mu^2 \delta \rho 
+ \rho k^{-2}{\cal V} \left[3 {\dot a \over a} 
(\mu^2 - w) + \frac{\dot w}{1+w} \right],
\end{equation}
where $\mu^2 =  F_{,X}/(F_{,X} + 2 X F_{,XX})$. If $K$ is a constant, then
$\mu^2=1$ so that the sound speed reduces to the manifestly stable $v^2=1$ on
small scales. Relaxing the assumption of a linear dependence of the Lagrangian
on $X$, different sound speed histories are possible. But let's see what
happens when we push this generalized scalar field across the $w=-1$ boundary.

Consider a dark energy model consisting of a single-component, generalized
scalar field. In order to cross at $w=-1$ we require $X=0$ and/or $F_{,X}=0$.
One can show that a vanishing $X$ corresponds to an extremum in the
equation-of-state, meaning $w=-1$ is a minimum or maximum.  This leaves
$F_{,X}=0$ as a necessary condition to cross. However, if $F_{,X}$ evolves
though zero and $F_{,XX}\neq 0$, then $\mu^2$ becomes negative, thereby leading
to unstable perturbations. Hence, the path across $w=-1$ would appear to be
blocked.

A remaining possibility is to allow for exceptional fine tuning of the scalar
field solution. Suppose that we fix $F$ such that $\mu^2 > 0$ for all
times. Then the scalar field equation of motion, 
\begin{equation}
\left[\mu^{-2} \ddot{\varphi} + 2 \frac{\dot a}{a} \dot{\varphi}\right]  
F_{,X} + a^2[F_{,\varphi} - 2 X F_{,X\varphi}] = 0,
\end{equation}
requires that we somehow tune $F_{,\varphi} - 2 X F_{,X\varphi} \to 0$ just as
$F_{,X}$ vanishes. Perhaps a solution can be constructed by working backwards,
starting from an assumed evolution for $w(\tau),\,\mu^2(\tau)$. But it should
be clear that even a slight shift away from this special trajectory through
phase space would prevent a crossing. The consequences of a perturbation at the
crossing are still worse, as we now argue, where terms like $\dot w /( 1 + w )$
indeed lead to unphysical behavior. We expand the scalar field relation
(\ref{eqn::presseqn}) and the perturbation equations (\ref{eqn::evolution}),
near the crossing time $\tau_\star$ in powers of $\tau-\tau_\star$. Hence, for
a trajectory, $w(\tau) = -1 + \alpha [\tau - \tau_\star]^n$ where $n$ is an odd
integer, we expand the fluid variables
\begin{eqnarray}
\delta &=& \delta^\star + \delta^1[\tau-\tau^\star]+... \cr
{\cal V} &=& {\cal V}^\star + ({\cal V}^1
+{\cal V}^L \ln|\tau-\tau^\star|)[\tau-\tau^\star]+... .
\end{eqnarray} 
The analytic solutions near the crossing give ${\cal V}^\star = 0$, and the
next coefficients can be easily obtained. The results show that
$\delta(\tau_\star + \epsilon) = \delta(\tau_\star - \epsilon)$ and ${\cal
V}(\tau_\star + \epsilon) = -{\cal V}(\tau_\star - \epsilon)$, for vanishing
$\epsilon$. Whereas the velocity perturbation vanishes at crossing, the density
perturbation need not. If $n=1$ then the pressure perturbation diverges
logarithmically, $\delta p \propto \ln|\tau - \tau^\star|$. If $n \ge 3$ then
the perturbations are well-behaved. But something else goes wrong in either
case.  Look at the expression for the density perturbation: $\delta\rho =
(F_{,\varphi} - 2 X F_{,X\varphi})\delta\varphi - (F_{,X} + 2 X F_{,XX})\delta
X$. If we require $\mu^2 \neq 0$ at crossing, then the coefficients of both the
$\delta\varphi$ and $\delta X$ terms must vanish; in order for the density to
be non-zero, one or both of $\delta\varphi$ and $\delta X$ must diverge.  If
the field and its derivatives are to make any sense, we must abandon this
scalar field description as a mechanism for crossing $w=-1$. Hence, there is no
viable path for the scalar field across the cosmological-constant boundary.
These findings are consistent with the thorough analysis of Vikman
\cite{Vikman:2004dc}.

If the scalar field itself cannot cross $w=-1$, then there is a simple way to
cross with two fields, as pointed out by Hu \cite{Hu:2004kh}.  Consider one
scalar field with equation-of-state $w_1 > -1$ and a second, generalized scalar
with $w_2 < -1$: together these can be used to describe a fluid with energy
density $\rho_{12} = \rho_1 + \rho_2$ dominated by the first field at early
times and the second field at late times. And since both components yield
stable fluctuations, the ensemble is also stable. Then, suppose we have a
trajectory $w_{12}(a)$ which describes such a dark energy that crosses at
$a=a_\star$. There is not a unique prescription to break this into two
components, but we can be economical by requiring that $w_1 \equiv \bar
w_1=$~constant for $a > a_\star$ and $w_2 = \bar w_2 =$~constant for $a <
a_\star$. Since $\rho_{12}(a) = \rho_{12}(a_0) \exp(3\int^{a_0}_a (1+w_{12}) d
\ln a)$, we can use energy conservation and the continuity of the energy
density to show that
\begin{eqnarray}
w_1(a < a_\star) &=& \bar w_2  + (w_{12}-\bar w_2) / \left[ 1 - {1 + \bar w_1 \over \bar w_1 - \bar w_2}
{\rho^\star_{12} \over \rho_{12}}
\left({a_\star  \over a}\right)^{3(1+\bar w_2)} \right] \cr\cr
w_2(a > a_\star) &=& \bar w_1  + (w_{12}-\bar w_1) / \left[ 1 - {1 + \bar w_2 \over \bar w_2 - \bar w_1}
{\rho^\star_{12} \over \rho_{12}}
\left({a_\star \over a}\right)^{3(1+\bar w_1)}\right] \cr\cr
f &=& 
\frac{ (1+\bar w_2)} { \bar w_2 - \bar w_1} 
\frac{ \rho_{12}^\star}{\rho_{12}^0}  
\left({a_\star \over a_0}\right) ^{3(1 + \bar w_1)}.
\label{weqns}
\end{eqnarray}
This last expression also gives the relative abundance of the first component
at the present day, $f = \rho_1^0 / \rho_{12}^0$. We have to be somewhat
careful with this piecewise construction, since $\dot w_{1,2}$ is discontinuous
at the crossing. If we choose $\bar w_{1,2}$ to be very close to $-1$, then
immediately before/after $a_\star$ the ratio  $\dot w_{1,2}/(1+w_{1,2})$ which
appears in (\ref{eqn::presseqn}) will be very large. Alternatively, two smooth
equation-of-state histories $w_1(a) \ge -1$ and $w_2(a) \le -1$ and a relative
abundance $f$ can be chosen to give a composite, scalar-field crossing model.

\begin{figure}
\includegraphics[scale=0.3]{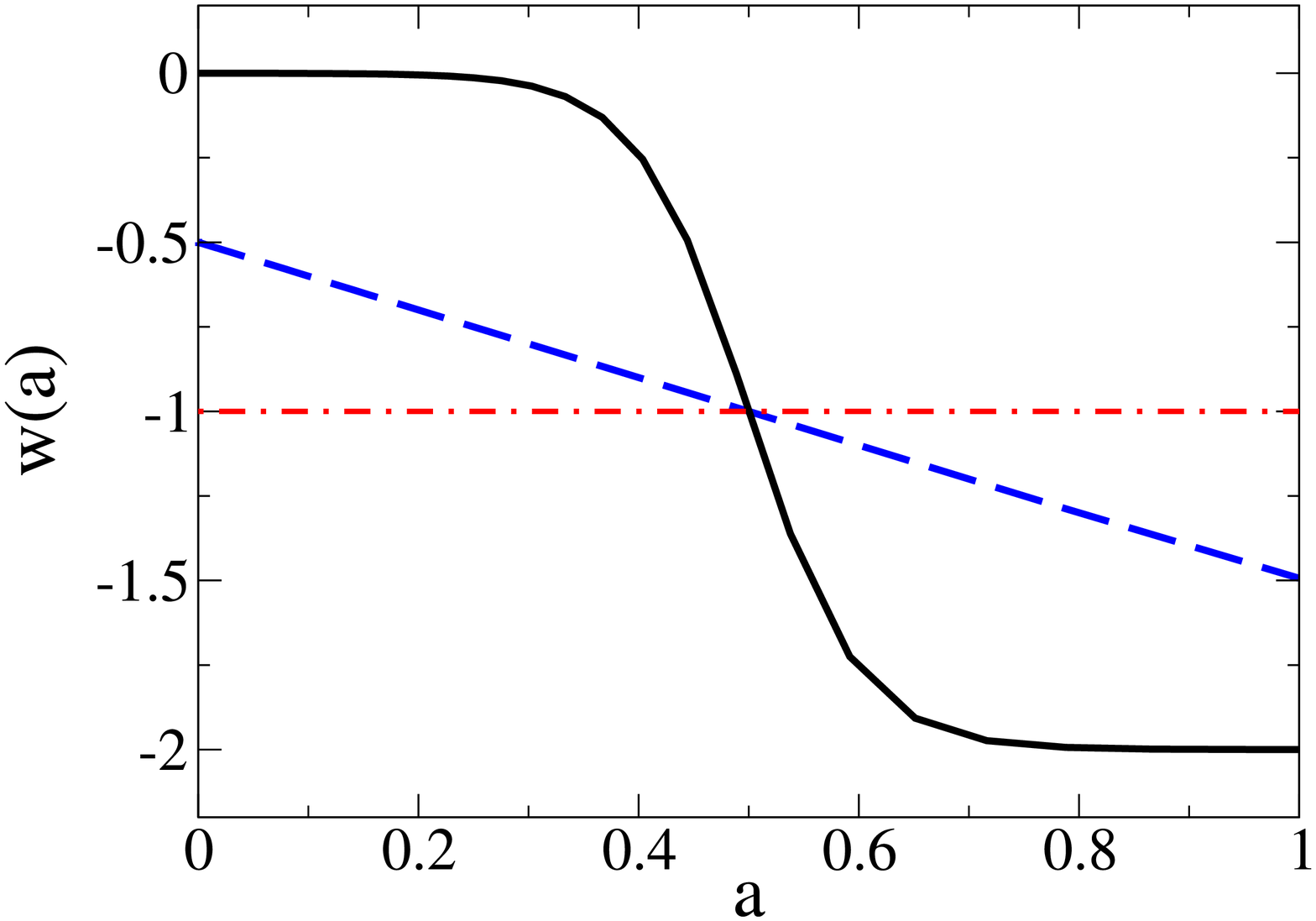}
\caption{
The equation-of-state $w(a)$ for two toy models which cross the
cosmological-constant boundary are shown. \protect{\eosa} is the dashed line,
and \protect{\eosb} is the solid curve. In both cases, dark energy evolves into
the phantom regime, below the dot-dashed line, beginning from $a_\star = 1/2$.
For \protect{\eosb} the dark energy evolves as matter at early times, since
$w\to 0$, and contributes measurably to the energy budget of the Universe
throughout matter domination.}
\label{fig::w}
\includegraphics[scale=0.3]{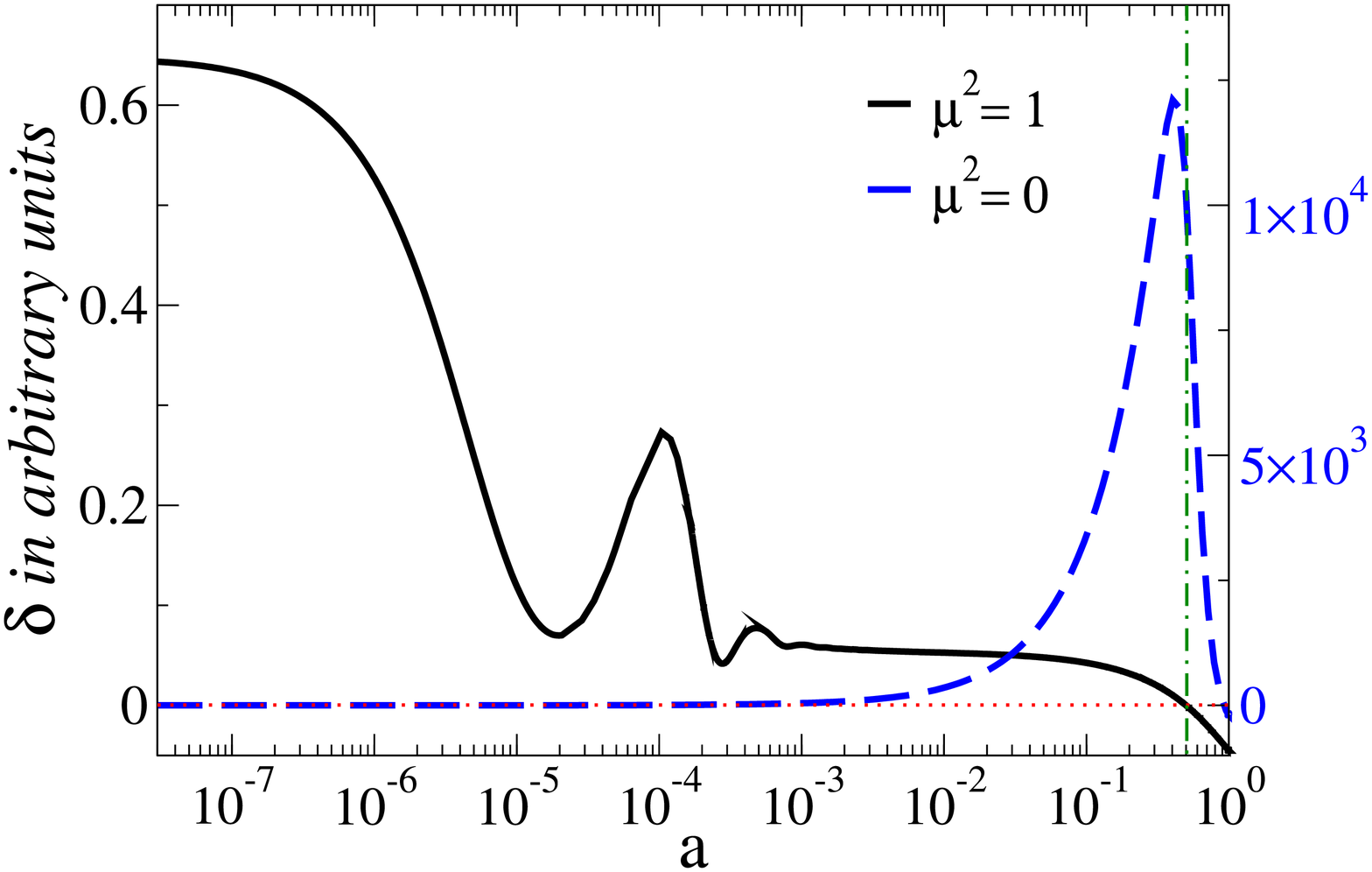}
\caption{
\label{fig::delta}
The density contrast for a pair of high-frequency modes with $\mu^2=1$ (solid
curve) and $\mu^2=0$ (dashed) for the \protect{\eosb} synthetic fluid. As
explained in the text, the density contrast for the $\mu^2=1$ case vanishes
when $w=-1$, instantaneously resembling a cosmological constant with no
fluctuations. In both cases the density contrast eventually grows negative, as
the gravitational instability flips to an antigravitational instability. (Note
the left scale is for the solid curve; the right scale is for the dashed
curve.)}
\end{figure}

To examine the observational consequences of a crossing dark energy component,
we consider two toy-model scenarios for $w(a)$. For the first (\eosa) we take
$w(a) = -3/2 + (1-a)$, and for the second (\eosb) we use $w(a) = -1 -
\tanh\left( 10\left [a-\frac{1}{2}\right ] \right)$. Both are shown in
Figure~\ref{fig::w}. Fixing $\Omega_m = 0.3$ today, then we find that \eosa\,
leads to negligible dark energy at early times whereas \eosb\, for which
$w\to0$ at early times, contributes a non-negligible fraction of the total
energy density throughout the matter-dominated era. Reducing the factor in the
tanh from 10, however, greatly reduces the abundance of dark energy in the
matte rera. We also investigated a third toy-model, consisting of a pair of
scalar fields with $w_1(a) \ge -1,\, w_2(a) \le -1$ which are smoothly-varying;
these components were contrived to produce the ensemble evolution of \eosb.
However, we found no relevant differences with the piecewise construction.

We modified {\sc Cmbeasy} \cite{Doran:2003sy} to study the consequences of
these models. In neither case is there a discernable trace of the mechanism
used to achieve a cosmological-constant crossing. The synthetic fluid and the
two-scalar approach produce essentially identical results. (We use the
log-likelihood statistic introduced in Ref.~\cite{Huey:1998se} to check for
degeneracies.) This holds whether $\mu^2=1$ or $v^2=1$ on all scales. There is
a distinction between models with $\mu^2=0$ and $\mu^2=1$, but this is
independent of the dark energy composition. Rather, perturbations in model
\eosb\, with $\mu^2=0$ grow from the initial, adiabatic conditions throughout
the matter-dominated era during which time the dark energy has $w=0$, too; in
the present era, the dark energy comes to dominate, $w$ crosses $-1$, and the
growth is not only slowed but reversed as the dark energy lumps respond
anti-gravitationally to the gravitational potentials. The observational imprint
is an additional ISW contribution and boost to the mass power spectrum, both
typically on the order of $\sim 10\%$ for the cases we have considered.
Reducing the factor in the tanh from 10 greatly reduces the impact of any
additional clustering.

The evolution of dark energy density perturbations in these models is of some
interest. For model \eosa\, the perturbations never grow large; the negative
equation-of-state and negligible energy density until very late times keep the 
fluctuations from making a significant impact. (To neglect the perturbations
completely, however, is equivalent to a violation of energy conservation. The
degree to which this influences theoretical predictions depends on the gauge in
which the fluctuations are ignored.) For \eosb\, the perturbations grow
significantly in the $\mu^2=0$ case. Starting from adiabatic initial
conditions, the density perturbations grow like $\delta \propto a$ through the
matter-dominated era. As illustrated in Figure~\ref{fig::delta}, when the dark
energy comes to dominate, the growth rate slows and eventually the density
contrast is driven negative. In contrast, the $v^2,\,\mu^2=1$ models suppress
fluctuation growth. We also notice that the density contrast for high-frequency
modes vanishes when $w$ crosses $-1$, as seen in Figure~\ref{fig::delta}.  For
these high frequency modes the oscillations have been long suppressed by the
expansion, and the density directly tracks the driving term, which is
proportional to $(1+w)$. When the sources vanish, so do the dark energy
fluctuations. Only in this instant does the dark energy instantaneously
resemble a smooth, cosmological constant. 

We have now described several simple ways to engineer a $w=-1$ crossing.  The
primary tool for modeling dark energy is the scalar field, which must possess
an unorthodox, negative kinetic term to achieve $w<-1$. There are substantial
reasons to find such a field objectionable \cite{Carroll:2003st}, although we
are willing to keep such possibilities open until observations and experiment
give us a better idea as to the nature of the dark energy. 

Other mechanisms have been proposed to explain a dark energy $w=-1$ crossing.
First, a dark matter-dark energy interaction  which siphons energy from CDM
into a quintessence field can produce an expansion history which  appears as
though the cosmos is dominated by a non-interacting CDM and a crossing dark
energy \cite{Wetterich:1994bg,Huey:2004qv,Amendola:1999er}. However, such a
coupling may be difficult to realize because of quantum effects
\cite{Doran:2002bc}. Likewise, scalar-tensor theories can mimic a crossing
under less-extreme circumstances than a negative-kinetic cosmic scalar field.
Second, a cosmic field which undergoes a burst of particle production has been
suggested as a means to produce pole-like inflation. Transplanting this
mechanism from early- to late-times, the field may be used to drive
super-acceleration without requiring a super-negative pressure
\cite{Barrow:2004xh,Zimdahl:2000zm}. Third, higher-order or non-perturbative
quantum effects as occur in the vacuum metamorphosis model can cause an
otherwise well-behaved scalar field to push the cosmos across the $w=-1$
boundary \cite{Parker:1999td}. In these scenarios there are other effects ---
variations in particle masses or coupling constants, features in the CMB
anisotropy and mass power spectra --- that may be exploited to identify the
underlying mechanism.  

If observational and experimental evidence grows sufficiently compelling to
favor a component that has $w < -1$ for some duration, and if conventional or
astrophysical effects cannot account for the observed phenomena, then it may be
necessary to consider the cosmological-constant boundary-crossing scenarios.


\begin{acknowledgments}

This work was supported in part by NSF AST-0349213 at Dartmouth and DFG grant
1056/6-3 at Heidelberg.

\end{acknowledgments}

\appendix



\begin{thebibliography}{}

\bibitem{Riess:2004nr}
A.~G.~Riess {\it et al.}  [Supernova Search Team Collaboration],
Astrophys.\ J.\  {\bf 607}, 665 (2004).

\bibitem{Spergel:2003cb}
D.~N.~Spergel {\it et al.}  [WMAP Collaboration],
Astrophys.\ J.\ Suppl.\  {\bf 148}, 175 (2003).

\bibitem{Readhead:2004gy}
A.~C.~S.~Readhead {\it et al.},
Astrophys.\ J.\  {\bf 609}, 498 (2004).

\bibitem{Goldstein:2002gf}
J.~H.~Goldstein {\it et al.},
Astrophys.\ J.\  {\bf 599}, 773 (2003)
[arXiv:astro-ph/0212517].

\bibitem{Rebolo:2004vp}
R.~Rebolo {\it et al.},
arXiv:astro-ph/0402466.

\bibitem{Tegmark:2003ud}
M.~Tegmark {\it et al.}  [SDSS Collaboration],
Phys.\ Rev.\ D {\bf 69}, 103501 (2004).

\bibitem{Hawkins:2002sg}
E.~Hawkins {\it et al.},
Mon.\ Not.\ Roy.\ Astron.\ Soc.\  {\bf 346}, 78 (2003).

\bibitem{Wetterich:fm}
C.~Wetterich,
Nucl.\ Phys.\ B {\bf 302}, 668 (1988).

\bibitem{Ratra:1987rm}
B.~Ratra and P.~J.~Peebles,
Phys.\ Rev.\ D {\bf{37}}, 3406  (1988).

\bibitem{Caldwell:1997ii}
R.~R.~Caldwell,~R.~Dave and P.~J.~Steinhardt,
Phys.\ Rev.\ Lett.\  {\bf 80}, 1582 (1998).

\bibitem{Caldwell:1999ew}
R.~R.~Caldwell,
Phys.\ Lett.\ B {\bf 545}, 23 (2002).

\bibitem{Vikman:2004dc}
A.~Vikman,
arXiv:astro-ph/0407107.

\bibitem{Hu:2004kh}
W.~Hu,
arXiv:astro-ph/0410680.

\bibitem{Ma:1995ey}
C.~P.~Ma and E.~Bertschinger,
Astrophys.\ J.\  {\bf 455} 7, (1995).
  
\bibitem{Armendariz-Picon:2000ah}
C.~Armendariz-Picon, V.~Mukhanov and P.~J.~Steinhardt,
Phys.\ Rev.\ D {\bf 63}, 103510 (2001).  
  
\bibitem{Erickson:2001bq}
J.~K.~Erickson, R.~R.~Caldwell, P.~J.~Steinhardt, C.~Armendariz-Picon and V.~Mukhanov,
Phys.\ Rev.\ Lett.\  {\bf 88}, 121301 (2002).

\bibitem{DeDeo:2003te}
S.~DeDeo, R.~R.~Caldwell and P.~J.~Steinhardt,
Phys.\ Rev.\ D {\bf 67}, 103509 (2003);
[Erratum-ibid.\ D {\bf 69}, 129902 (2004)].

\bibitem{Bean:2003fb}
R.~Bean and O.~Dore,
Phys.\ Rev.\ D {\bf 69}, 083503 (2004).

\bibitem{Weller:2003hw}
J.~Weller and A.~M.~Lewis,
Mon.\ Not.\ Roy.\ Astron.\ Soc.\  {\bf 346}, 987 (2003).

\bibitem{Press:1980}
W.~H.~Press and E.~T.~Vishniac,
Astrophys.\ J.\ {\bf 239}, 1 (1980).
 
\bibitem{Doran:2003sy}
M.~Doran,
arXiv:astro-ph/0302138.

\bibitem{Huey:1998se}
G.~Huey, L.~M.~Wang, R.~Dave, R.~R.~Caldwell and P.~J.~Steinhardt,
Phys.\ Rev.\ D {\bf 59}, 063005 (1999).
 
\bibitem{Carroll:2003st}
S.~M.~Carroll, M.~Hoffman and M.~Trodden,
Phys.\ Rev.\ D {\bf 68}, 023509 (2003).

\bibitem{Wetterich:1994bg}
C.~Wetterich,
Astron.\ Astrophys.\  {\bf 301}, 321 (1995).

\bibitem{Amendola:1999er}
L.~Amendola,
Phys.\ Rev.\ D {\bf 62}, 043511 (2000).

\bibitem{Huey:2004qv}
G.~Huey and B.~D.~Wandelt,
arXiv:astro-ph/0407196.

\bibitem{Doran:2002bc}
M.~Doran and J.~Jaeckel,
Phys.\ Rev.\ D {\bf 66}, 043519 (2002).

\bibitem{Barrow:2004xh}
J.~D.~Barrow,
Class.\ Quant.\ Grav.\  {\bf 21}, L79 (2004).

\bibitem{Zimdahl:2000zm}
W.~Zimdahl, D.~J.~Schwarz, A.~B.~Balakin and D.~Pavon,
Phys.\ Rev.\ D {\bf 64}, 063501 (2001).
 
\bibitem{Parker:1999td}
L.~Parker and A.~Raval,
Phys.\ Rev.\ D {\bf 60}, 063512 (1999);
[Erratum-ibid.\ D {\bf 67} (2003) 029901].

\end{thebibliography}
\end{document}